\documentclass
[aps,prc,twocolumn,showpacs,showkeys,amsmath,floatfix,superscriptaddress]
{revtex4-1}
\usepackage{graphicx}
\usepackage{mathptmx}                
\usepackage{dcolumn}                 
\usepackage{bm}                      
\usepackage{sistyle} 			
\usepackage{eurosym} 			
\DeclareMathOperator{\e}{e}
\newcommand{\be}{\begin{equation}}
\newcommand{\ee}{\end{equation}}
\newcommand{\bea}{\begin{eqnarray}}
\newcommand{\eea}{\end{eqnarray}}

\usepackage{color}
\usepackage{soul,latexsym, amssymb, amsmath, multirow, mathrsfs, booktabs}
\usepackage{longtable}
\usepackage{rotating}

\begin{document}

\title{In-medium nuclear  cluster energies within the
Extended Thomas-Fermi approach}

\author{Fran\c{c}ois Aymard}
\affiliation{CNRS and ENSICAEN, UMR6534, LPC, 14050 Caen c\'edex, France}
\author{Francesca Gulminelli}
\affiliation{CNRS and ENSICAEN, UMR6534, LPC, 14050 Caen c\'edex, France}
\author{J\'er\^ome Margueron}
\affiliation{Institut de Physique Nucl\'eaire de Lyon, Universit\'e Claude Bernard Lyon 1, \\IN2P3-CNRS, F-69622 Villeurbanne Cedex, France}

\begin{abstract}
A recently introduced analytical model for the nuclear density profile~\cite{pana} is implemented in 
the Extended Thomas-Fermi (ETF) energy density functional.
 This allows to (i) shed a new light on the issue of the sign of surface symmetry energy in  nuclear mass formulas, 
which is    strongly related to the non-uniformity of the isospin asymmetry in finite nuclei, as well as to (ii) evaluate the 
in-medium corrections to the nuclear  cluster energies in thermodynamic conditions relevant for the description 
of the (proto)-neutron star crust.
The  ground state configurations of the model are compared to Hartree-Fock calculations in spherical symmetry for some selected isotopic chains, and systematic 
errors are quantified. 
The in-medium modification of the nuclear mass due to the presence of  a gas component is shown to strongly
depend both on the density and the asymmetry of the nucleon gas. This shows the importance of accounting for such effects in the
realistic modelizations of the equation of state for core-collapse supernovae and proto-neutron stars. 
\end{abstract}

\date{\today}

  

\maketitle

\section{Introduction}

The semi-classical Thomas Fermi (TF) and Extended-Thomas-Fermi (ETF) approach to the density functional theory were largely 
used in the 80's for nuclear structure applications~\cite{Brack1985,krivine,Centelles1990} as well as astrophysical ones~\cite{Suraud1984,Pi1986}. 
Two motivations of searching for approximations of the microscopic mean-field theory
with effective interactions can be advanced. On one side, this semi-classical quasi-analytical theory provides a clear physical insight  on the functional dependence of nuclear energies and density profiles which cannot be achieved with the numerical resolution of HF equations for single-particle orbitals.
On the other side, the computational ressources at that time made systematic HF calculations very hard to perform with reliable numerical error 
bars. The exponential progress of numerical computing in the next two decades made this motivation obsolete and we have assisted to an impressive progress of mean-field and beyond-mean field large-scale nuclear structure calculations~\cite{beyond-mf}. 
However in the recent years, a renewed interest towards the ETF theory has appeared ~\cite{etf_recent,chamel_etf,mekjian}. 
This is largely due to the new challenges which are open to the field  and the needs for a microscopic description 
of the very exotic nuclear species which are expected to exist in stellar matter.  
The  widely used equation of state models for  supernova matter, neutron stars, and proto-neutron star typically replace the nuclear distribution in stellar matter with a single representative nucleus, and use  density functionals to describe the nucleus as well as the surrounding nucleon gas~\cite{LS,Shen}.
 More recent models have replaced the single nucleus approximation by a statistical distribution of nuclei, 
using the  experimental information for nuclear masses and  , doing so, do not consider in-medium modifications to the nuclear cluster energies~\cite{hempel,mishustin,japanese,us}. 
A complete and microscopic description of stellar matter at finite temperature and at
sub-saturation densities implies the evaluation of an extremely large data base of ground states and excited nuclear 
configurations in a dilute light-particles environment  which are not directly
accessible to variational HF calculations, or, for some of them which are accessible, which are computationally too 
expensive for large-scale calculations~\cite{newton_thesis}.

In this context it is interesting to develop an ETF based formalism which would, in a   quasi-analytical way, 
provide nuclear clusters energies for ground state and excited state configurations
using  energy functionals optimized for exotic nuclear data as well as   neutron matter 
calculations~\cite{sly4}.
In a recent paper~\cite{pana}, we have proposed a model based on a simple  parameterization of
Fermi-Dirac density profiles and on the zeroth order TF approximation for the kinetic energies and currents.
 
 Comparing this model to HF ground-state configurations, a good agreement was reached since the
differences between the model and the HF calculation were found  independent of the gas density and of the order of 
0.5-1~MeV/nucleon.
The model has therefore been employed to evaluate the in-medium energy shifts 
in a large variety of excited state configurations~\cite{pana,esym}. 

In this paper we introduce second order $\hbar^2$-corrections, allowing the introduction of the spin-orbit interaction and 
an increased precision in the evaluation of the kinetic energy density.
 The agreement with HF energies is therefore found to be better.
This improved model is used to obtain the functional form of the nuclear energies as a function of mass number and 
asymmetry, both in the case of ordinary nuclei in the vacuum, and in the case of nuclei immersed in a nucleon gas.

The paper is organized as follows:
In Section \ref{sec:model} we recall the ETF formalism and present two possible modelling of the nuclear density profiles 
 employed in the variational ETF. 
These two parametrizations are critically compared in Section  \ref{sec:HF}, which presents a comparison of the 
ETF model to HF calculations in order to assess the 
 accuracy of the ETF calculation. 
It is shown that the inclusion of second order $\hbar^2$-terms considerably improves the predictive power of the model.
Section  \ref{sec:sym_ene} presents an application of the model to the study of the functional dependence 
of the symmetry energy on the nuclear mass.
We  analyze  the well known problem of the sign of the surface symmetry energy~\cite{Myers1985}, and show 
that an explicit comparison to HF calculations can help to eliminate the ambiguity in the decomposition between surface and bulk in 
the two-component nuclear system. 
Section  \ref{sec:gas} reports detailed results concerning the modification to the nuclear energy due to the presence of 
 a gas component. 
The specific case of a nucleus immersed in a neutron gas, 
similar to the ground state of nuclear clusters present in the crust of neutron stars,
is examined.
The case where the nucleus is in an arbitrary single-particle excited state configuration, as it is the case in the finite temperature
conditions of supernova matter and proto-neutron stars, is also considered and shown to lead to very different energy shifts.

\section{The model}\label{sec:model}

 We briefly present the model for nuclei and nuclear matter which is based on the Skyrme interaction~\cite{sly4} and 
where the semi-classical ETF approximation is employed.
This approximation requires a parametrization of the nuclear density profiles and two types of such parametrizations
are investigated and compared.

\subsection{Skyrme functionals and ETF semi-classical approximation} \label{sec:ETF}

The Skyrme functional for the time-even energy density is expressed as~\cite{Bender2003a,Bartel2008}
\begin{eqnarray}
\mathcal{E}_{sky}(r) &=& \frac{\hbar^2}{2m}\tau^{(0)}+\sum_{t=0,1} C_t^\rho(\rho^{(0)}) \rho^{(t)2} + C_t^{\Delta \rho} \rho^{(t)} 
\Delta \rho^{(t)} \nonumber \\
&+& C_t^\tau \rho^{(t)}\tau^{(t)}
 +\frac{1}{2}C_t^J J^{(t)2}+C_t^{\nabla J} \rho^{(t)}\nabla\cdot J^{(t)},
\label{eq:functionalt}
\end{eqnarray}
where the superscripts $t=0$ and $t=1$ stand for the isoscalar and isovector part of the corresponding densities, as for example,
\begin{eqnarray}
\rho^{(0)}(r) &=& \rho_n(r)+\rho_p(r),\nonumber \\
\rho^{(1)}(r) &=& \rho_n(r)-\rho_p(r).
\end{eqnarray}
The coefficients $C$ are taken to be constants except for $C_t^\rho$ which depends of the isoscalar density
$\rho^{(0)}$ according to the parameterization,
\begin{equation}
C_t^\rho(\rho^{(0)}) = C_t^\rho(0) + (C_t^\rho(\rho_{sat})-C_t^\rho(0))\left(\frac{\rho^{(0)}}{\rho_{sat}}\right)^\alpha,
\end{equation}
where $\rho_{sat}$ is the saturation density in infinite symmetric nuclear matter.
See Appendix \ref{coefficients} and Ref.~\cite{Bender2003a} for additional definitions.

This functional depends on the occupied single particle orbitals in a complex way because of the presence of 
kinetic densities and currents. A simpler dependence on the single-particle densities $\rho_q$ can be obtained
using a semi-classical Wigner-Kirkwood expansion~\cite{Brack1985}, which is the basis of the so-called Thomas-Fermi approximation.
We will consider an expansion up to the second $\hbar^2$-order. The kinetic density $\tau^{(0)}$ reads
at the zeroth order (Thomas Fermi approximation)~\cite{Brack1985}
\be
\tau^{(0)}=\tau_{TF}\equiv \frac{3}{5}(3\pi^2)^{(2/3)} \sum_q \rho_q^{5/3}, \label{eq:TF}
\ee
and at the second order $\tau^{(0)}=\tau_{TF}+\sum_q \tau_{q,2}^L+\tau_{q,2}^{NL}$
where,
\begin{eqnarray}
\tau_{q,2}^{L}&=&\frac{1}{36}\frac{(\nabla\rho_q)^2}{\rho_q}+\frac{1}{3}\Delta\rho_q, \label{tau2L}\\
\tau_{q,2}^{NL}&=&\frac{1}{6}\frac{\nabla\rho_q\cdot\nabla f_q}{f_q}
+\frac{1}{6}\rho_q\frac{\Delta f_q}{f_q}
-\frac{1}{12}\rho_q\left(\frac{\nabla f_q}{f_q}\right)^2\nonumber\\
&&+\frac{1}{2}\left(\frac{2m}{\hbar^2}\right)^2\rho_q\left(\frac{W_q}{f_q}\right)^2. \label{tau2NL}
\end{eqnarray}
Here,  $\tau_{q,2}^{L}$ is the second order local term, 
$\tau_{q,2}^{NL}$ is the second order non-local term, and the effective mass factor $f_q=m/m_q^*$ is defined as
\begin{equation}
f_q = 1 + \frac{2m}{\hbar^2}[(C_0^\tau+C_1^\tau)\rho_q+ (C_0^\tau-C_1^\tau)\rho_{\bar q} ].  \label{eq:mstar}
\end{equation}
.
 
The spin-orbit current obtained at the same $\hbar^2$-order in the semiclassical expansion
is given by~\cite{Brack1985}
\begin{equation}
J_q = -\frac{2m}{\hbar^2 f_q} \rho_q W_q,
\label{eq:socurrent}
\end{equation}
where the spin-orbit potential $W_q$ reads
\begin{eqnarray}
W_q &=& -(C_0^{\nabla J}+C_1^{\nabla J})\nabla\rho_q
-(C_0^{\nabla J}-C_1^{\nabla J})\nabla\rho_{\bar q} \nonumber \\
&&+(C_0^{J}+C_1^{J})J_q+(C_0^{J}-C_1^{J})J_{\bar q}. 
\label{eq:sopotential}
\end{eqnarray}

In asymmetric systems, the relation between the spin currents $J_n$ and $J_p$ and
the gradient of the densities is given by the solution of the 2x2 system of linear equations~\cite{Bartel2008},
\begin{eqnarray}
&&\left(\frac{\hbar^2}{2m}f_q+(C_0^J+C_1^J)\rho_q\right)J_q + (C_0^J-C_1^J)\rho_qJ_{\bar q} \nonumber \\
&&=(C_0^{\nabla J}+C_1^{\nabla J})\rho_q\nabla\rho_q+(C_0^{\nabla J}-C_1^{\nabla J})\rho_q\nabla\rho_{\bar q}.
\end{eqnarray}
The solutions $J_n$ and $J_p$ of this system are injected in Eq.~ (\ref{eq:socurrent}) in order to obtain
the expression of $W_q$ in terms of the density gradients.

Let us notice however that in several Skyrme interactions such as SIII, SLy4, SGII,...  the terms
in $J^{(t)2}$ in the functional~ (\ref{eq:functionalt}) are neglected.
The spin-orbit potential $W_q$ given in Eq.~ (\ref{eq:sopotential}) is therefore simply
related to the gradient densities in these functionals, and we have
\begin{eqnarray}
W_q = -(C_0^{\nabla J}+C_1^{\nabla J})\nabla\rho_q
-(C_0^{\nabla J}-C_1^{\nabla J})\nabla\rho_{\bar q}. 
\label{eq:approxW}
\end{eqnarray}

In principle, fourth order $\hbar^4$-terms can also be added for an improved predictive power, 
as it has already been done in previous works \cite{Brack1985,krivine,Centelles1990}.

Hereafter the Skyrme functional with the kinetic energies and currents approximated within the second order
 ETF expansion 
will be noted $\mathcal{E}_{sky}^{ETF}[\rho^{(0)},\rho^{(1)}]$.

\subsection{Symmetric nuclei and generalized Fermi function solution}\label{sec:GFD}

The great advantage of the semi-classical ETF approximation is that the 
non-local terms in the energy density functional, see Eqs.~(\ref{tau2L}), (\ref{tau2NL}) and (\ref{eq:sopotential}),
are entirely replaced by local gradients.
As a consequence, the energy functional solely depends on the local particle densities. 
Thus, the energy of any arbitrary nuclear configuration can be calculated if the density profiles $\rho_q$ are 
 given through a parametrized form. 
 
 The ground state configuration should in principle be obtained from the variational calculation, which in the single density
case, $\rho=\rho^{(0)}$ and $\rho^{(1)}=0$, is reduced to single Euler-Lagrange equation,
\begin{equation}
\frac{\partial\mathcal{E}_{sky}^{ETF}}{\partial\rho}
-\nabla\cdot \frac{\partial\mathcal{E}_{sky}^{ETF}}{\partial\nabla\rho}
+\Delta \frac{\partial\mathcal{E}_{sky}^{ETF}}{\partial\Delta\rho} = \lambda, \label{eq:Euler}
\end{equation}
where $\lambda$ is a Lagrange multiplier imposing the correct particle number.
 The generalization of Eq.~(\ref{eq:Euler}) to the two-density case realized in isospin-asymmetric nuclei
is straightforward~\cite{Brack1985}.

Substituting Eq.~ (\ref{eq:functionalt}) into the Euler-Lagrange equation, and using the $\hbar^2$-order in the ETF expansion 
Eqs. (\ref{tau2L}),  (\ref{tau2NL}), (\ref{eq:mstar}),  (\ref{eq:approxW}),  leads to
\begin{equation}
\lambda=\frac{dh}{d\rho}+   C^\nabla  \left(  {\nabla}\rho\right)^2
+C^\Delta \Delta\rho
, \label{eq:Euler_1D}
\end{equation}
with
\bea
C^\nabla(\rho)&=&\frac{\hbar^2}{2m}\frac{1}{36} \left(  \frac{1}{\rho^2}+ \frac{3\kappa^2}{f^2}\right)  
+\frac{C_0^{\nabla J}B_J}{2f^2}  , \nonumber \\
C^\Delta(\rho)&=&\frac{\hbar^2}{2m}\frac{1}{3} \left(  -\frac{1}{6\rho}  +  \frac{7}{3}\kappa   -  \frac{\kappa}{2f} \right) -2 C_0^{\Delta\rho}+C_0^{\nabla J}B_J\frac{\rho}{f}  , \nonumber \\
 h(\rho) &=&
\frac{\hbar^2}{2m}  f\tau_{TF} +\rho^2 C_0^\rho ,
\label{eq_sym_deriv_h}
\eea
where the following quantities have been introduced: $\kappa=2m/\hbar^2C_0^\tau$, $B_J=2m/\hbar^2C_0^{\nabla J}$,
$f=1+\kappa\rho$.
This equation was solved, within a simplified energy functional and in the semi-infinite slab geometry, in Ref.~\cite{krivine}. 
 A numerical solution of the Euler-Lagrange equations for finite nuclei, employing more general density functionals,
including the Coulomb interaction and possibly $\hbar^4$-terms in the semi-classical expansion, is as numerical demanding 
as the resolution of the HF equations. 
For this reason, trial density profiles containing only a few variational 
parameters are often employed \cite{Brack1985,krivine,Centelles1990,etf_recent,chamel_etf,mekjian}.
In particular,
in Ref.~\cite{krivine} it was shown that a trial density  presenting the correct asymptotic behaviors in the one-dimensional system, 
is given by the
generalized Fermi-Dirac distribution (GFD):
\begin{equation}
\rho_{GFD}(r) \equiv \frac{\rho_{sat}}{ ( 1+\exp (r-R_\nu)/a_\nu)^{\nu}}.
\label{eq:GFD}
\end{equation}
 The parameter $\rho_{sat}$ coincides with the solution of the Euler-Lagrange equation in the
limit of infinitely extended nuclei, which is the saturation density of symmetric nuclear matter.
The other parameters $a_\nu$, $R_\nu$ and $\nu$ are analytically derived from
the asymptotic solution of the Euler-Lagrange equation~\cite{krivine}. 
Details are given in Appendix  \ref{analytic}. 

For the nuclear interactions considered in this work, the terms in $J^{(t)2}$ in the functional~ (\ref{eq:functionalt}) are 
neglected. The correction to the Euler equation~(\ref{eq:Euler_1D}) induced by the inclusion of the spin-orbit current is 
given in Appendix \ref{bj}.

\subsection{Simple Fermi function  model}\label{sec:FD}

The variational approach presented in Section  \ref{sec:GFD} allows an analytical determination of the nuclear energy for symmetric $N=Z$ nuclei.
Unfortunately, the generalization of these equations to asymmetric nuclei is highly non-trivial~\cite{krivine_iso} 
unless severe approximations are assumed. Since our aim is to have a robust model which can be applied for exotic nuclei as well as for dilute nuclear clusters present in the (proto)neutron star crust, we shall not consider uncontrolled approximations.
 We shall therefore propose a modified functional 
form, which is inspired by the solution of the Euler-Lagrange equations, and is directly optimized on Hartree-Fock calculations.

In Ref.~\cite{pana} an analytical modelling of the density profile was proposed, using a simple $\nu=1$ Fermi-Dirac (FD) 
functional form.
 \begin{equation}
\rho_{FD}(r) \equiv \frac{\rho_{sat}(\delta)}{  1+\exp (r-R)/a}.
\label{eq:FD}
\end{equation}
 Similarly to the previous model given by Eq.~(\ref{eq:GFD}), the parameter $\rho_{sat}(\delta)$ matches with the limit of
infinitely large nuclei, but this time can be generalized in asymmetric matter. 
The asymmetry dependence of the saturation density is well given by the following form~\cite{pana},
\begin{equation}
\rho_{sat}(\delta) = \rho_{sat} \left( 1-\frac{3 L_{sym} \delta^2}{K_{sat}+K_{sym} \delta^2 }  \right),
\label{paperpana1:eq:rho0}
\end{equation}
In this expression, $L_{sym}=3\rho_{sat}\partial \mathcal{E}_{sky}^{sym}/\partial \rho^{(0)}$ 
and $K_{sym}=9\rho_{sat}^2\partial^2 \mathcal{E}_{sky}^{sym}/\partial \rho^{(0)2}$ 
are the slope and curvature of the symmetry energy at saturation, where 
we have introduced the usual definition of the symmetry energy functional:
\be
\mathcal{E}_{sky}^{sym}(\rho^{(0)} )= \frac{1}{2}\rho^{(0)2} \frac{\partial^2 \mathcal{E}_{sky}^{ETF}}{\partial 
\rho^{(1)2}}|_{\rho^{(1)}=0}.
\ee
In Eq.~(\ref{paperpana1:eq:rho0}),  the parameter $\delta=1-2\rho_{sat,p}/\rho_{sat}$ is the bulk asymmetry 
of the nucleus. 
The bulk asymmetry differs from the global asymmetry of the nucleus, $I=1-2Z/A$, because of the 
presence of a neutron skin and  Coulomb effects.
The relation between the bulk asymmetry $\delta$ and the global asymmetry $I$
is given by~\cite{ldm,centelles1,centelles2}:
\begin{equation}
\delta = \frac{I+\frac{3 a_C}{8 Q} \frac{Z^2}{A^{5/3}}} {1+\frac{9 E_{sym}}{4 Q} \frac{1}{A^{1/3}}},
\label{paperpana1:eq:deltacl}
\end{equation}
where $E_{sym}=\mathcal{E}_{sky}^{sym}[\rho_{sat}]/\rho_{sat}$ is the symmetry energy at saturation,  
$Q$ is the surface stiffness coefficient extracted from a semi-infinite nuclear matter calculation
and $a_C$ is the Coulomb
parameter taken equal to $a_C=0.69$ MeV.
The radius parameter $R$ entering the density profile~(\ref{eq:FD}) is given by
\begin{equation} 
R = R_{HS} \left[ 1 - \frac{\pi^2}{3} \left(\frac{a}{R_{HS}}\right)^2 \right],
\label{paperpana1:eq:radiusws}
\end{equation} 
where $R_{HS} = (3A/4\pi\rho_{sat}(\delta))^{1/3}$   is the equivalent homogeneous sphere radius. 
Eq.~(\ref{paperpana1:eq:radiusws}) can be deduced from the general expression given in the appendix 
Eq.~(\ref{eq_sym_3D_R_nu}).
The diffuseness parameter $a$ of the  total density profile is assumed to depend quadratically on $\delta$, 
$a=\alpha+\beta\delta^2$, where $\alpha$ and $\beta$ were fitted from HF calculations in Ref. ~\cite{pana}.

This simple model has the great advantage that its level of predictivity is the same 
for symmetric and asymmetric nuclei ~\cite{pana}.
To describe an isospin asymmetric system, we need two independent   density profiles.
We will for this purpose use the total isoscalar density $\rho$ and the proton density $\rho_p$ as two FD functions
characterized by the corresponding saturation densities $\rho_{sat}$, and $\rho_{sat,p}=(1-\delta)/2\rho_{sat}$, 
diffuseness $a(\delta)$, $a_p(\delta)$ (with parameters $\alpha$, $\alpha_p$, $\beta$, $\beta_p$ 
governing the diffuseness isospin dependence),
and radii $R,R_p$ as given by Eq.~(\ref{paperpana1:eq:radiusws}) above.

Moreover, the extension to the physical situation of the inner crust, where nuclei are immersed in a gas of continuum states,
is also relatively straightforward~\cite{pana}. This point will be discussed in Section  \ref{sec:gas}.

\section{Comparison to Hartree-Fock calculations} \label{sec:HF}

Once the parameters of the  density profiles are specified, the nuclear ground state energy is straightforwardly calculated as
\be
E_{ETF}=\int d^3r \mathcal{E}_{sky}^{ETF}\left [\rho^{(0)},\rho^{(1)}\right ], \label{eq:ETF}
\ee
where the semi-classical ETF nuclear functional $\mathcal{E}_{sky}^{ETF}$ is given in sec.~\ref{sec:ETF}.
In this expression,  the isoscalar and isovector densities are calculated
imposing FD profiles for the total isoscalar and proton densities  $\rho$ and $\rho_p$. 
For symmetric nuclei the isovector density $\rho^{(1)}$ vanishes and it becomes possible to describe 
the density profile with a GFD profile via Eq.~(\ref{eq:GFD}), which in principle should be 
more correct since it corresponds to the variational solution of the ETF problem, 
 though using a trial density profile.

The quality  of the models  given by Eqs.~(\ref{eq:GFD}) and (\ref{eq:FD})  can be judged 
by comparing the ansatz density profiles and the associated energies to HF calculations  performed 
with the same nuclear effective interaction. 
For these numerical applications, we will systematically use the SLy4 Skyrme nuclear interaction~\cite{sly4}.
We first  compare the GFD~(\ref{eq:GFD})  and FD~(\ref{eq:FD}) ansatz density profiles for $N=Z$ nuclei, showing the minor
role of the parameter $\nu$  as well as the limitations of the variational approach.

\subsection{Comparison between FD and GFD ETF in symmetric nuclei}

\begin{figure}[tb]
\includegraphics[width=\linewidth,clip]{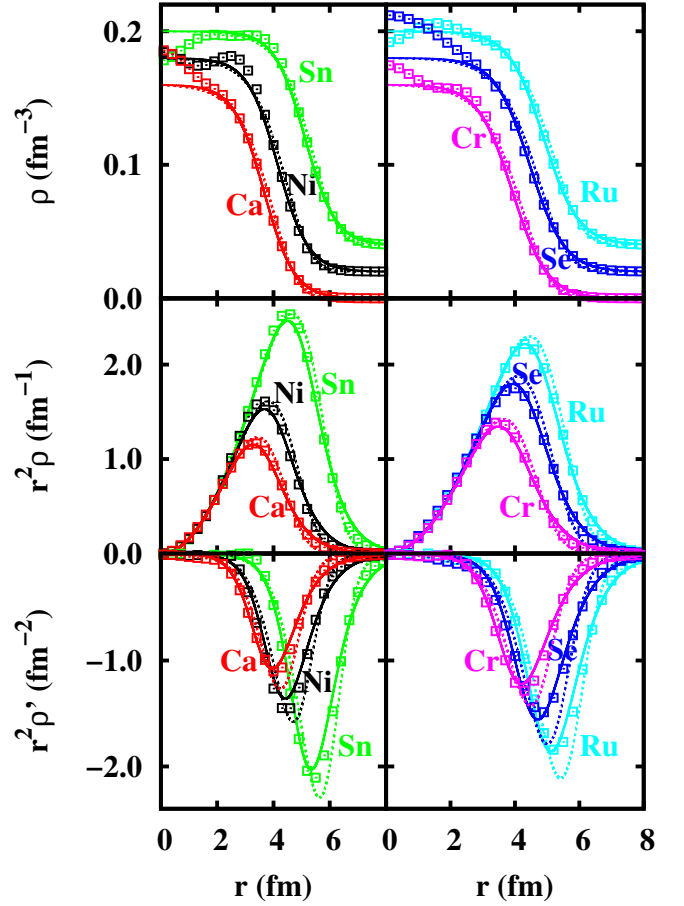}
\caption{(Color online) Density profiles (upper panel), corresponding particle numbers (central panel) 
and density derivative profile times $r^2$ (lower panel)
of different magic (left side, $^{40}$Ca, $^{56}$Ni and $^{100}$Sn) 
and open-shell (right side, $^{48}$Cr, $^{68}$Se, $^{88}$Ru) even-even symmetric nuclei. 
Symbols: spherical HF calculations. Dashed lines: GFD model Eq.~(\ref{eq:GFD}). 
Full lines: FD model Eq.~(\ref{eq:FD}). A vertical shift of $\delta\rho=0.02(0.04)$ fm$^{-3}$ is applied to the 
density profiles of $^{56}$Ni, $^{68}$Se ($^{100}$Sn, $^{88}$Ru)  to better separate the different curves.}
\label{fig:profiles_A}
\end{figure}

Figure  \ref{fig:profiles_A} shows the density profiles, as well as the density multiplied by 
  $r^2$  and the gradient of the density $\times r^2$
for some chosen representative $N=Z$ nuclei. 
In all cases, the GFD~(\ref{eq:FD}) and  FD~(\ref{eq:GFD}) ansatz density profiles are compared to Hartree-Fock calculation 
in spherical symmetry. 
Double magic nuclei are considered in the left part of the figure, while open shells ones are plotted in the right part.

\begin{figure}[tb]
\includegraphics[angle=270,width=\linewidth,clip]{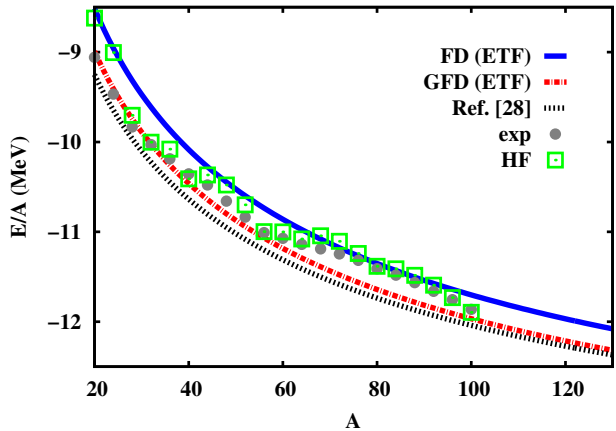}
\caption{(Color online) Energy per nucleon of $N=Z$ nuclei as a function of the mass number. squares: HF calculations. circles: experimental data from ~\cite{nudat}. Full blue line: FD model. Dashed red line: GFD model. Dotted black line: results from ref.~\cite{dan}. }
\label{fig:energy_A}
\end{figure}

We can see that both the FD and the GFD  
ansatz can reproduce the HF density profiles with the same accuracy 
 and, quite interestingly, the diffuseness of the nuclear surface is equally well reproduced by the two ansatz.
Microscopic density profiles exhibit ripples in the central density which are not accessible to a simple (G)FD shape.
However, these structures are not expected to influence the energetics of the system in an important way, because of the volume
element in the energy integral. 
Moreover, it is known that they are to a large extent artefacts of the mean-field approach and are expected to be washed out by correlations.
 For these reasons, the densities and the gradient of the densities are multiplied by $r^2$ in the lower panels of 
Fig.~\ref{fig:profiles_A}.
Interesting enough, the GFD functional form, even multiplied by $r^2$, does not give a better reproduction of the microscopic 
calculations than the simpler FD one. 
It is clear from this figure that the FD profile is flexible enough to reproduce the gross features of the microscopic calculation.  
 In particular we can see that the fall-off of the density in the  HF calculation is very well described by an exponential behavior.
Conversely, it was shown in Ref.~\cite{Centelles1990} that the
variational ETF solution exhibits a slower polynomial decrease when the $\hbar^4$-terms are included.
This is again an argument suggesting that
we can safely neglect these higher order terms. 
It is also important to remark that the Coulomb interaction is known to affect the density profile, though it can be considered as a second order effect. 
The Coulomb effects are implicitly included in the FD model of section \ref{sec:FD}, while 
both the direct and exchange term of the Coulomb energy density  should be included
in the Euler-Lagrange equations for a correct   derivation of the density profile  if we use the variational strategy 
of section \ref{sec:GFD}.  

The satisfactory performance of the FD model is confirmed and quantified by Fig.~\ref{fig:energy_A}, which displays the energy per particle of $N=Z$ nuclei as a function of their mass number.  
Only the nuclear part of the HF energy is considered in this figure. 
For consistency, the same Coulomb energy, as obtained in HF, is subtracted from the experimental nuclear masses, taken from~\cite{nudat}.

\begin{figure}[tb]
\includegraphics[angle=270,width=\linewidth,clip]{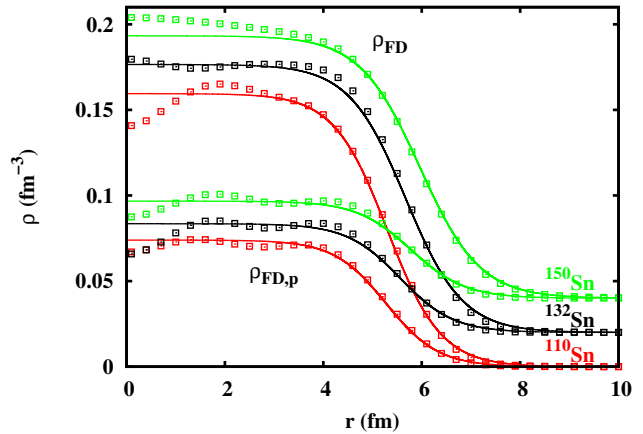}
\caption{(Color online)  Total (upper curves) and proton (lower curves) density profiles of different Sn isotopes. 
 Symbols: spherical HF calculations.  
Full lines: FD model Eq.~(\ref{eq:FD}).  A vertical shift of $\delta\rho=0.02(0.04)$ fm$^{-3}$ is applied to the density profiles of 
$^{132}$Sn ($^{150}$Sn),  to better separate the different curves. }
\label{fig:profiles_NZ}
\end{figure}

We can see that the GFD profiles systematically produce more binding than the FD ones, as expected from the 
wider variational space associated to this functional form. 
The resulting energies are in good agreement with both the microscopically calculated and measured masses for 
the magic nuclei represented in Fig.~\ref{fig:profiles_A} above. 
However, the other nuclei are overbound.
 This overbinding is known to be due to the absence of fourth-order terms in the ETF functional~\cite{Centelles1990}. 
The simpler FD model with no variationally determined parameter underbinds magic nuclei, 
but it leads to an overall good agreement with the 
microscopic calculations. 
 These results are consistent with previous findings comparing FD and GDF ansatz profiles~\cite{Centelles1990}.
 
A liquid-drop-like parametrization for the nuclear masses in the framework of mean-field Skyrme models 
was recently proposed in Ref.~\cite{dan}. 
In this reference, the authors propose the following functional form for the nuclear energy
\begin{equation}
E_{LDM}=a_v A -a_s A^{2/3}-\frac{a_v^a}{1+\frac{a_v^a}{a_s^a A^{1/3}}} A I^2, \label{dan}
\end{equation}
and  have extracted the parameters $a_v$, $a_s$, $a_v^a$, $a_s^a$ from a fit of HF calculations in an uncharged semi-infinite 
geometry,  as well as from the neutron-proton radii differences.
 The isoscalar part of Eq.~(\ref{dan}) contains bulk and surface contributions only while the isovector part contains additionally
curvature and beyond  contributions.
The result of Eq.~(\ref{dan}), using the same SLy4 functional~\cite{sly4}, is also displayed in Fig.~\ref{fig:energy_A}.
We can see that the variational ETF calculation correctly converges towards the slab estimation~(\ref{dan}) for very large mass numbers, where curvature corrections to the surface energies due to the spherical geometry  are becoming negligible. 
 The functional form given by Eq.~(\ref{eq:ETF}) naturally contains curvature effects, in the isoscalar and isovector channel.
The difference between Eq.~(\ref{eq:ETF}) and Eq.~(\ref{dan}) is mostly due to the missing curvature term in the isoscalar channel
in Eq.~(\ref{dan}).
For light nuclei Eq.~(\ref{dan}) therefore tends to overestimate the binding.
 
From the ensemble of results presented in Fig.~\ref{fig:energy_A} we can conclude 
that  the ansatz densities FD and GFD reproduces equally well the microscopic HF calculations, and that the
biggest source of discrepancy is mainly due to the lack of shell effects in the ETF approach. 
We therefore stick to the FD parametrization, and turn to test its predictivity in asymmetric nuclei, 
where a direct analytical solution of the Euler-Lagrange equations does not exist  with any trial density profile. 

 \subsection{Comparison between HF and FD ETF in asymmetric nuclei} \label{sec:HF-asym}

\begin{figure}[tb]
\includegraphics[angle=270,width=\linewidth,clip]{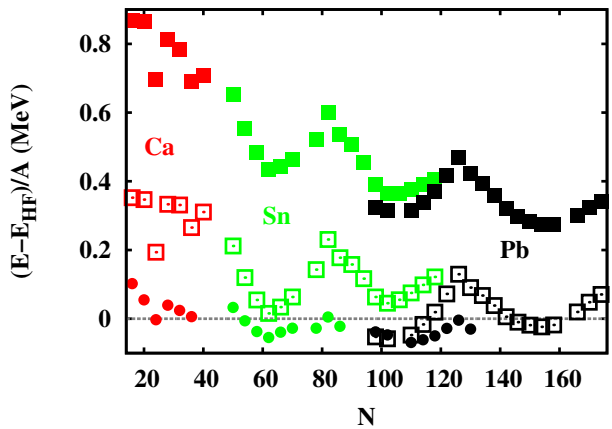}
\caption{(Color online) Difference between the energy per particle calculated in the ETF model and in the HF
for the isotopic chain of Ca (red symbols), Sn (green symbols), and Pb (black symbols). Full squares: zeroth order TF approximation. 
Empty squares: second order $\hbar$ expansion. Full circles: experimental data.  }
\label{fig:energy_NZ}
\end{figure}

Some representative microscopic HF density profiles  are compared to the FD ansatz~(\ref{eq:FD}) in Fig.~\ref{fig:profiles_NZ}. 
We can see that the level of agreement with the microscopic calculation is comparable to the case of symmetric nuclei.
It does not depend on the exoticity of the nucleus but mostly on the size of the system.  
The larger the system, the better the FD model.
This statement is better quantified in Fig.~\ref{fig:energy_NZ}, which shows the energy difference between the ETF calculation and the microscopic one as a function of the neutron number, for some selected isotopic chains. 
In this figure, the filled symbols correspond to the Thomas-Fermi or local density approximation, consisting in truncating the kinetic 
energy density expansion to the zero order in $\hbar$, see  Eq.~(\ref{eq:TF}). 
In this approximation, which  was used in a previous work~\cite{pana}, the spin-orbit term vanishes
and the local kinetic energy density at a position $r$ is the same as for infinite nuclear matter at the local density 
$\rho(r)$, $\rho_p(r)$.
We can see that the inclusion of second order terms in the functional (open symbols in Fig.~\ref{fig:profiles_NZ})
considerably improves the description.
In particular, for the heaviest isotopic chain considered, the average ETF energy very well reproduces the average HF energy.
The deviations  are comparable to the difference between the HF model and the experimental data (full circles), and 
can be fully ascribed to the missing shell effects.  These effects, which cannot be accounted by a semiclassical model as ETF, could in principle be included with Strutinsky smoothing techniques~\cite{etf_recent}. For the application to the proto-neutron star crust that we are interested in, we however do not expect this to be an important point, as shell effects are known to rapidly wash out with increasing temperature. 

To conclude,  the use of the simple FD ansatz in the
ETF approach at second order in $\hbar$  has been found to reproduce with a good accuracy the microscopic HF density
profiles as well as the HF binding energies, with an accuracy
of the order of 300 keV/nucleon for the lighter nuclei, and which does not exceed 150 keV/nucleon for the heavy ones. 
 A significant improvement is found with respect to the previous work~\cite{pana}.

\section{Symmetry energy from ETF} \label{sec:sym_ene}

Let us now turn to a first application of the model.
Given the reasonably good reproduction of the smooth part of the microscopic nuclear density, the ETF description can be used 
to explore the functional form of the nuclear mass, and in particular the separation in a bulk and surface term of its isovector and isoscalar parts.

\begin{figure}[tb]
\includegraphics[angle=270,width=\linewidth,clip]{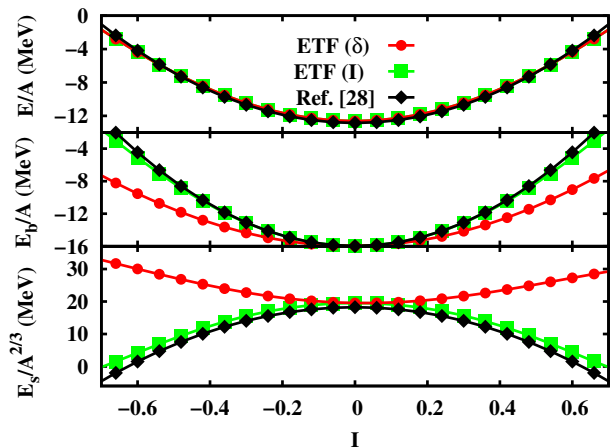}
\caption{(Color online)  Total (upper panel) and bulk (central panel)  
energy per nucleon and surface energy per surface nucleon (lower panel)
along the isobaric chain $A=200$. 
Red circles: ETF calculation including the neutron skin effect Eq.~(\ref{esurf})
(see text).
Green squares: ETF calculation neglecting the neutron skin effect Eq.~(\ref{esurf_I}).  
Black diamonds: estimation from Eq.~(\ref{esurf_dan}). }
\label{fig:esym}
\end{figure}

Such a separation is important for the extraction of the largely unknow density behavior of the symmetry energy 
from nuclear data~\cite{esym_book}. 
Indeed it has been proposed in the literature~\cite{frdm} that the symmetry energy can be strongly constrained from the measurement 
of nuclear masses. 
These estimations give the experimental constraints on the symmetry energy which have at present the smallest 
uncertainties~\cite{tsang}. 
The determination of the symmetry energy from nuclear mass implies that the surface and bulk component
of the isospin dependence can be unambiguously distinguished. 
However, very different values are reported in the literature for the surface symmetry energy coefficient~\cite{dan03,reinhardt,pei,douchin}.
 In a two-component system, 
there are two possible definitions of the surface energy which depend on the definition of
the bulk energy in the cluster~\cite{Myers1985,Farine1986,Centelles1998}:
the first one corresponds to identifying the bulk energy of a system of $N$ neutrons and $Z$ protons to the energy of an equivalent
piece of nuclear matter $E_{S}=S_{\gamma_e}\equiv E-eA$, where $A=N+Z$ and $e$ is the energy per nucleon of uniform matter. 
The second definition $E_{S}=S_{\gamma_\mu}\equiv E-\mu_n N-\mu_p Z +pV$ corresponds to the grandcanonical thermodynamical Gibbs definition.
The first definition is the standard surface energy of the droplet model~\cite{droplet}, while 
 the second one gives the quantity to be minimized in the variational calculation 
 conserving proton and neutron number. 
It was shown that the sign of the surface symmetry energy depends on the choice between these two 
possibilities~\cite{Myers1985,Farine1986,Centelles1998}. 
Moreover it was argued~\cite{Myers1985} that the case of liquid drop model (LDM) mass formulas, 
where the bulk energy is a function of the total mass number $A$ and of the global asymmetry $I=(N-Z)/A$ only, 
is closer (though not equal) to the Gibbs definition. 
This can explain why LDM mass formulas systematically obtain negative (though widely varying)
surface symmetry energy coefficients ~\cite{dan03,reinhardt,pei,douchin}.

 If the total energy $E$ is exactly known, the two decompositions
are in principle exactly equivalent, meaning that the surface symmetry energy is ill-defined.
However, the total energy is never exactly known. In the case of empirical mass formulas, it is given by a fit of experimental data.
In the case of ETF based functionals, as in the present study, we are seeking for the best possible approximation to the complete 
variational HF problem within a given effective interaction.
Therefore it is important to determine  if there is a decomposition which is best suited to reproduce
the Hartree-Fock energy.
The variational ETF theory imposes the use of local quantities instead of global ones, and it therefore naturally leads to the use of the local asymmetry parameter $\delta$
instead of $I$. This choice implies that the surface symmetry energy shall be positive as we will show hereafter.

\subsection{Surface symmetry energy} \label{surf1}

 In our model, the neutron and proton density profiles are fully defined by the
FD ansatz, which conserves the particle numbers by construction
and contains no variational parameters. For this reason  we do not need to introduce the Gibbs surface tension~\cite{Myers1985}, 
and will only refer to the definition of  the surface energy  as the quantity  deduced from the total energy after subtraction 
of the energy the system would have in the absence of the surface:
\be
E_S= E_{ETF}
- \frac{\mathcal{E}_{sky}^{ETF}\left [\rho^{(0)}=\rho_{sat}(\delta),\rho^{(1)}=\rho_{sat}(\delta)\delta \right ]}{\rho_{sat}(\delta)} A .
 \label{esurf}
\ee

As we have already observed, because of the presence of the neutron skin, the isospin asymmetry distribution is 
not uniform in the nuclear  system. 
As a consequence, the bulk asymmetry $\delta$ does not coincide with the  global asymmetry $I=(N-Z)/A$, 
see Eq.~(\ref{paperpana1:eq:deltacl}). 
 It is clear that the symmetry energy obtained from Eq.~(\ref{esurf}) will be different if one replace the subtracted bulk
component evaluated at the bulk asymmetry $\delta$ by the one evaluated at the global symmetry $I$.
Most mass formulas, both phenomenological~\cite{ldm,frdm} and microscopically motivated ~\cite{dan03,reinhardt,pei,douchin}, 
assume  however that the bulk isospin dependence is given by the the global asymmetry variable $I=(N-Z)/A$. 
This is for instance the case of  the reported Eq.~(\ref{dan}) where the surface energy is defined as~\cite{dan},
\be
E_S^{LDM}= E_{LDM}- \left ( a_v  - a_v^a I^2 \right ) A  \label{esurf_dan} .
\ee
 In Fig.~\ref{fig:esym} are compared, as a function of the global asymmetry $I$, the energy and the symmetry energy 
obtained from Eqs.~(\ref{eq:ETF}) and (\ref{esurf}), referred to as ETF($\delta$), the same energies but replacing $\delta$ 
by $I$, 
\be
E'_S=E_{ETF} 
- \frac{\mathcal{E}_{sky}^{ETF}\left [\rho_{sat}(I),\rho_{sat}(I) I \right ]}{\rho_{sat}(I)} A , \label{esurf_I}
\ee
referred to as ETF(I), and the ones obtained from Eqs.~(\ref{dan}) and (\ref{esurf_dan}), referred to as Ref.\cite{dan}. 
This comparison is performed
for a representative isobaric chain $A=200$. 
For such heavy nuclei, 
 the curvature terms play a minor role and the liquid-drop formula (\ref{esurf_dan}) referred as Ref.\cite{dan}
leads to a nuclear energy very close to the ETF model.
However, because of the very different partition between bulk and surface in the models EFT($\delta$) and LDM (\ref{esurf_dan}),  
the surface symmetry energy shows an opposite behavior in the two models. 
As a consequence,  the surface energy, and more specifically the surface symmetry energy, depends on the prescription employed
to remove the bulk component, cf Eqs.~(\ref{esurf}) and (\ref{esurf_I}).

 It is interesting to notice the very close behavior of the surface energies given by ETF(I) and LDM (\ref{esurf_dan}) in 
Fig.~\ref{fig:esym}. 
This very similar behavior assets the important role of the asymmetry parameters $\delta$ and $I$.
Specifically, the isospin dependence of the symmetry energy shown also in Fig.~\ref{fig:esym} is
found the behave in a opposite way between the models EFT($\delta$) and the two other models ETF(I) and LDM (\ref{esurf_dan}).
Consistently with Ref.~\cite{Myers1985}, it can be deduced from the curvature  of the curves represented in the bottom 
panel of Fig.~\ref{fig:esym} that the choice of the asymmetry variable have an important consequence on the sign of the
surface symmetry energy.

This effect is easy to understand analytically. 
Let us start from the relation between the bulk asymmetry $\delta$ and the global asymmetry $I$ previously given by Eq.~(\ref{paperpana1:eq:deltacl}). 
In the limit of small asymmetries, neglecting the Coulomb correction and fixing $x=3a_C/8Q$ and $y=9 E_{sym}/4Q$, 
we can  make the approximation
\be
\delta^2= \left (\frac{I+ x A^{1/3}(1-I)^2}
 {1+ y A^{-1/3}} \right )^2 \approx
I^2 \left (1-2y A^{-1/3} + g(A,I) \right), \label{delta_approx}
\ee
where the residual term $g(A,I)$ contains terms of order $x, y^2$ or smaller, which can 
be viewed as a correction with respect to the previous term. 
 We can see that  the replacement of the asymmetry
parameter $\delta$ by $I$  in eq.~(\ref{delta_approx}),  induces a correction to the LDM which is proportional to $A^{-1/3}$.
It means that the ambiguity in defining the proper asymmetry parameter in the bulk term of the
LDM propagates to the surface term.
Moreover, replacing $\delta$ by $I$ in the LDM induce an extra surface symmetry term with a negative sign, 
cf Eq.~(\ref{delta_approx}).
Since the surface symmetry term is positive in ETF($\delta$), the change of sign in ETF(I) can be
related to the negative extra term in Eq.~(\ref{delta_approx}).
In order to set this argument straight, let us now be more quantitative.

In the parabolic approximation~\cite{pana}, the bulk part of the ETF energy~(\ref{eq:ETF}) is quadratic in $\delta$:
\be
E_{ETF}\approx \left ( \lambda_{sat}  + E_{sym} \delta^2 \right ) A + E_S(A,I),\label{parabolic}
\ee
where $\lambda_{sat}=\mathcal{E}_{sky}^{ETF}(\rho_{sat},0)/\rho_{sat}$, see Eq.~(\ref{eq_,ym_lambda_nucl_matter})
in the appendix, and $E_S$ is the surface energy as in Eq.~(\ref{esurf}), $\lim_{A\to\infty} E_S/A=0$. 
If the same parabolic approximation is employed for the bulk term of Eq.~(\ref{esurf_I}) as it is customary done, see Eq.~(\ref{esurf_dan}),
\be
E_{ETF}\approx \left ( \lambda_{sat}  + E_{sym} I^2 \right ) A + E^\prime_S(A,I) \label{parabolic2} .
\ee
Comparing Eqs.~(\ref{parabolic}) and (\ref{parabolic2}), and using Eq.~(\ref{delta_approx}), we immediately get the
following relation between the two surface energies,
\be
E^\prime_S \approx E_S - \left (\frac{9 E_{sym}^2} {2Q}  A^{2/3} - Ag(A,I) \right ) I^2. \label{es_delta_I} 
\ee
 It is interesting to remark that this same equation was derived in Ref.~\cite{Myers1985} as the difference between 
the microcanonical ($\gamma_e$) and grandcanonical ($\gamma_\mu$) surface energies, in the limit of small asymmetries.
This equation shows that the surface energy $E_S'$ contains an extra negative symmetry term due to the  non-uniformity of the isospin distribution. 
As a result, the surface symmetry energy can change from positive to negative, as it is shown in Fig.~\ref{fig:esym}.

\subsection{Curvature symmetry energy} \label{surf2}
\begin{figure}[tb]
\includegraphics[angle=270,width=\linewidth,clip]{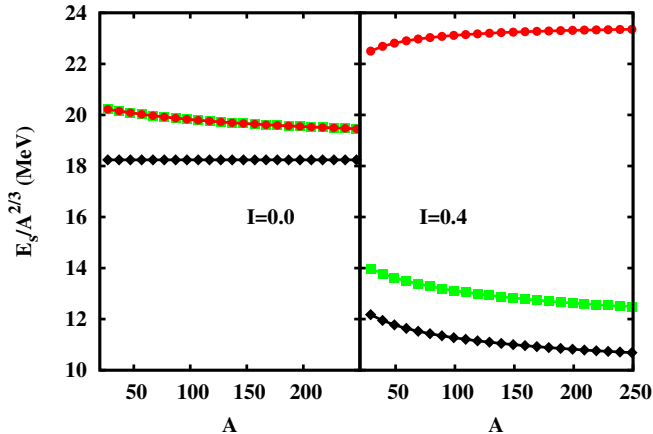}
\caption{(Color online) 
Surface energy per surface nucleon as a function of the nucleus mass. 
Red circles: ETF calculation including the neutron skin effect Eq.~(\ref{esurf}) (see text).
Green squares: ETF calculation neglecting the neutron skin effect Eq.~(\ref{esurf_I}). 
Black diamonds: estimation from Eq.~(\ref{esurf_dan}).
}
\label{fig:curv}
\end{figure}

In spherical symmetry it is well known that the surface energy obtained from Eq.~(\ref{esurf}) 
does not exactly scale as $A^{2/3}$, but it contains slower varying terms, the dominant one being a 
curvature term, proportional to $A^{1/3}$. 
 In Fig.~\ref{fig:curv} is displayed the behavior with the mass number $A$ of the surface energy
divided by $A^{2/3}$, where the surface energy is obtained in various ways:
the red circles represent the surface energy defined by Eq.~(\ref{esurf}), where the bulk asymmetry 
parameter $\delta$ is employed taking into account the non-uniformity of the isospin asymmetry
distribution in nuclei due to the presence of a neutron skin ; the green squares are obtained from 
Eq.~(\ref{esurf_I}), where the bulk asymmetry is approximated by the global asymmetry parameter $I$ ;
and the black diamonds stand for surface energy deduced from the LDM~(\ref{esurf_dan}).
The left panel of Fig.~\ref{fig:curv} shows the isoscalar behavior of the surface energy, where the
global asymmetry is fixed to be $I=0$, while the right panel shows the result by fixing the
asymmetry parameter to a finite value $I=0.4$.
Apart from the LDM~(\ref{esurf_dan}) in the isoscalar case, it is observed that the surface energy
is not constant, revealing the presence of a curvature energy in the considered models.

We can see from the left panel of Fig.~\ref{fig:curv} that the isoscalar curvature energy is positive for the ETF models
(\ref{esurf}) and (\ref{esurf_I}), and zero for the  LDM~(\ref{esurf_dan}).
The absence of the curvature energy in the isoscalar part of the functional~(\ref{esurf_dan}) is due to the fact that 
this LDM formula was motivated by one-dimensional slab calculations~\cite{dan} which by definition do not contain 
this term. 
The absence of a curvature energy is at the origin of the poorer reproduction of nuclear masses for symmetric 
nuclei, as observed in Figure  \ref{fig:energy_A}.  
 
 For the isoscalar case, there is almost no difference between the asymmetry parameters $I$ and $\delta$,
therefore the surface energies (\ref{esurf}) and (\ref{esurf_I}) overlap on the left panel of Fig.~\ref{fig:curv}.
On the right panel of Fig.~\ref{fig:curv} where $I=0.4$ the symmetry energy (\ref{esurf}) is shifted up, and
the symmetry energies (\ref{esurf_I}) and (\ref{esurf_dan}) are shifted down, as expected from Fig.~\ref{fig:esym}.
The curvature in the case $I=0.4$ is however given by a mixture of isoscalar and isovector contributions.
The effect of the isovector term in the case of the surface energy~(\ref{esurf}) is however sufficiently
negative to overcome the isoscalar contribution.
We can therefore deduce from Fig.~\ref{fig:curv} that the curvature energy is positive and the asymmetry curvature 
energy is negative in the case of Eq.~(\ref{esurf}).
In the case of the model (\ref{esurf_I}), the isovector term goes in the same direction as the isoscalar term, and the
trend of the surface symmetry is similar to the one from the LDM~(\ref{esurf_dan}).
 We can see from Figure~\ref{fig:curv} that again the sign of the surface symmetry term is opposite in the ETF models
(\ref{esurf}) and (\ref{esurf_I}), and that the ETF model neglecting the neutron skin effect (\ref{esurf_I}) have the
same behavior as the mass formula~(\ref{esurf_dan}).

 This is again coming from the bulk contribution subtracted in the two ETF models~(\ref{esurf}) and ~(\ref{esurf_I}).
Indeed Eq.~(\ref{es_delta_I}) shows that the difference between bulk $\delta$ and global $I$ isospin parameters 
induces an extra mass dependent term, which contributes negatively to the curvature surface symmetry energy.
 Neglecting the non-uniform isospin density distribution, induced by neutron skin and Coulomb repulsion, a positive
symmetry curvature energy is obtained, while taking into account the non-uniformity of the isospin density distribution,
a negative sign is found.

\subsection{Hints from Hartree-Fock}

According to the discussion in Sections  \ref{surf1}, \ref{surf2}, an ambiguity exists in the definition and in  the sign of the surface 
symmetry energy, as well as of the curvature symmetry energy. 
This ambiguity arises from the fact that the bulk asymmetry of nuclei $\delta$ differs from their global asymmetry $I$
because of the presence of a neutron skin and, to a minor extent, to the distortion of the density profile due to the 
Coulomb interaction. 
Since $I=\delta$ at the thermodynamic bulk limit, a priori both Eq.~(\ref{esurf}) and Eq.~(\ref{esurf_I})
can be proposed as a definition of the surface energy, and one may conclude that the surface symmetry energy is  ill-defined.

At the level of the ETF approximation however, these two equations are not equivalent and only Eq.~(\ref{esurf}) 
is theoretically  justified. 
Indeed, as we have discussed in Section  \ref{sec:GFD},  
 if we consider only ground state configurations,
the ETF approximation is equivalent to the solution of a set of coupled local Euler-Lagrange equations. In the idealized situation of a system with a locally constant density profile
($\rho_q'(r)=\rho_q''(r)=0$ for a given value of $r=r_0$), these equations simply read
\be
\lambda_q=\frac{\partial h}{\partial \rho_q}(r_0) .
\ee
This equation  admits the simple local bulk solution $\rho_q(r_0)=\rho_{sat,q}$, where the saturation density $\rho_{sat,q}$
has to be calculated at the asymmetry $\delta(r_0)=1-2\rho_p(r_0)/\rho(r_0)$, that is the local 
asymmetry. This reasoning implies that the bulk energy has to be calculated with the local bulk asymmetry $\delta$.

Another argument going in the same direction comes from a comparison to HF calculations.
Indeed, for the quantity defined in Eq.~(\ref{esurf_I}) to vanish at the bulk limit, 
the $\rho_{sat,q}$ parameters entering the  proton and neutron density profiles should be identified
with  $\rho_{sat,q}=(1\pm I)/2\rho_{sat}(I)$, and the one entering the total isoscalar density should read $\rho_{sat}=\rho_{sat}(I)$.
Replacing these quantities in Eq.~(\ref{eq:FD}) leads to a different model both for the density profiles and for the ETF energy
according to Eq.~(\ref{eq:ETF}). This alternative model, noted ETF(I) to distinguish it from the ETF($\delta$) 
proposed in Section  \ref{sec:FD}, can be compared to HF calculations using the same Skyrme functional.
This comparison in shown in table  \ref{table} for the representative case of the total energy per nucleon 
and proton mean radius along the Pb isotopic chain. 
 
\begin{table}[tb]
\caption{ Comparison between HF and ETF along the Pb isotopic chain. The different columns give, from left to right:
the mass number of the isotope, the average isospin asymmetry, the bulk isospin asymmetry 
(Eq.~(\ref{paperpana1:eq:deltacl})), the mean HF proton radius, the percentage deviation in the mean proton radius between HF 
and the ETF(I) and the ETF($\delta$) models, the HF energy per nucleon,  the percentage deviation in the energy per nucleon between HF and the ETF(I) and the ETF($\delta$) model. The last line gives the arithmetic average along the isotopic chain.}
\begin{ruledtabular}
\begin{tabular}{c|cccccccc}
A & 
$I$ & $\delta$ &  
$\overline{r}_p(HF)$ & $\%(I)$ & $\%(\delta)$ & 
$E_{HF}/A$ & $\%(I)$ & $\%(\delta)$\\
\hline
$180$ & 0.09 & 0.08 & 5.31 & -0.5 & -1.1 & -12.25 & -0.3 & -0.4 \\
$196$ & 0.16 & 0.13 & 5.40 & +0.8 & -0.5 & -11.93 & +0.4 & -0.1 \\
$216$ & 0.24 & 0.19 & 5.50 & +2.7 & +0.4 & -11.34 & +1.8 & +0.5 \\ 
$236$ & 0.31 & 0.25 & 5.67 & +3.4 & +0.0 & -10.51 & +1.8 & -0.2 \\
$256$ & 0.36 & 0.29 & 5.77 & +5.0 & +0.6 &  -9.81 & +3.6 & +0.7 \\
\hline
$\langle \rangle$  & 0.23 & 0.19 & 5.53 & +2.3 & -0.1 & -11.17 & +1.4 & +0.1
\end{tabular}
\end{ruledtabular}
\label{table}
\end{table}

We can see that the ETF($\delta$) model systematically gives a better reproduction of HF results, and the deviation 
between ETF($\delta$) and ETF(I) increases with increasing difference between bulk $\delta$ and global $I$ asymmetry
parameters. 
The HF result supports the intuitive idea behind Eq.~(\ref{eq:FD}) which is related to
the local character of the Euler-Lagrange variational equations: the density in the bulk of a heavy nucleus is related 
to the saturation density corresponding to the local bulk asymmetry, and not to the global
asymmetry of the nucleus.

 In conclusion, these two arguments shows that the model EFT($\delta$) is better justified both from a
theoretical point of view and from a comparison to HF calculations.
The bulk energy shall therefore be parameterized in terms of the bulk asymmetry, and the surface symmetry
energy in the corresponding LDM shall be positive.


\section{Nuclei immersed in a nucleon gas} \label{sec:gas}

We now turn to the second application of this model, which concerns the evaluation of the in-medium modification
of the nuclear ground-state energy due to the presence of a surrounding nuclear gas  of unbound nucleons.
Having in mind the evaluation of the equation of state and structure of   supernova matter~\cite{esym},
we have to consider  excited states of arbitrarily high energy. Above the particle separation threshold, vibrations and deformations 
can be neglected and the excited configurations essentially correspond  to the  coexistence of nuclei of arbitrary isospin with 
 a uniform neutron gas composed both of protons and neutrons. 
The extension of the formalism presented in Section~\ref{sec:FD} to this situation was already presented in Ref.~\cite{pana}.
Here we give only the main points of the model, and address the reader to Ref.~\cite{pana} for further details.
In a Wigner-Seitz cell occupied by a uniform nucleon gas with densities $\rho_{g,q}$, the total density profile of protons and neutrons 
in the cell can be decomposed into a cluster and a gas component.
 Due to the high nuclear incompressibility, we assume that the bulk density of the clusters is not modified by the occupation
of unbound particle states~\cite{pana}.
Eq.~(\ref{eq:FD}) is then replaced by a more general ansatz, including the uniform gas, and given by
 \begin{equation}
\rho_{FD,q}(r) \equiv \frac{\rho_{sat,q}(\delta)-\rho_{g,q}}{ 1+\exp (r-R_q)/a_q}+\rho_{g,q}.
\label{eq:FD-gas}
\end{equation}
The bulk asymmetry $\delta$ of the cluster has also to be modified from the vacuum expression 
Eq.~(\ref{paperpana1:eq:deltacl}) in order to include the overlap of the cluster with the uniform gas. 
Indeed Eq.~(\ref{paperpana1:eq:deltacl}), being an equation for a bound nucleus,  applies only to the bound 
part of the cluster $A_e$.
In the spirit of the independent particle model, this bound part can be defined as the ensemble of bound states, 
obtained from the total number of particles with the subtraction of the gas contribution,
\be
A_e=[1-\rho_g/\rho_{sat}(\delta)] A  \; \; ; \;  Z_e=[1-\rho_{g,p}/\rho_{sat,p}(\delta)] Z
\ee
where $\rho_g=\rho_{g,n}+\rho_{g,p}$ is the total isoscalar gas density.

The bulk asymmetry $\delta_e$ of the ensemble of bound cluster states  is given by Eq.~(\ref{paperpana1:eq:deltacl})
with $A=A_e$, $Z=Z_e$ and $I=1-2Z_e/A_e$, while the local asymmetry in the bulk of the cluster is estimated as a 
linear combination of the asymmetries coming from the bound and the unbound components:
\begin{equation}
\delta = \left ( 1-\frac{\rho_g}{\rho_{sat}(\delta)}\right ) \delta_e + \frac{\rho_g}{\rho_{sat}(\delta)} \delta_g,
\label{eq:deltar}
\end{equation}
where $\delta_g=1-2\rho_{g,p}/\rho_g$ is the gas asymmetry. 

 The total energy in the presence of a gas is still given by Eq.~(\ref{eq:ETF}), but it now depends both
on the cluster and on the gas density profiles through Eq.~(\ref{eq:FD-gas}):
\bea
E^{tot}_{ETF}&=&\int  \mathcal{E}_{sky}^{ETF}\left [\rho_{FD,n}+\rho_{FD,p},\rho_{FD,n}-\rho_{FD,p}\right ] d^3r \\
&=&E^{tot}_{ETF}(A,\delta,\rho_g,\delta_g). \nonumber
, \label{eq:ETF_gas}
\eea
For an application to the equation of state at finite temperature for supernovae matter~\cite{hempel,mishustin,japanese,us},
all the possible values of  $A,\delta,\rho_g$, and $\delta_g$ have to be considered, and the relative weight of the different 
configurations is given by the Boltzmann factor. 
If we limit ourselves to a  neutron rich nuclear cluster embedded in a pure neutron gas, 
the quality of the model  can again be judged in comparison to HF calculations.
In Ref.~\cite{pana}, it was shown that the quality of reproduction of complete HF results of this model is 
almost independent of the presence of
an external gas.

The presence of a nucleon gas obviously modifies the energy of the nuclear cluster. 
The in-medium modification of the cluster energy $\delta E_m$ can be computed by subtracting 
to the total energy the contribution of the gas alone and of the nucleus alone, according to~\cite{pana,esym}:
\be
\delta E_{m}=E_{ETF}^{tot}-E_{ETF}(A,Z)-V_{WS} \mathcal{E}_{sky}^{ETF}\left [\rho_g,\rho_g\delta_g\right ],
\label{total_energy_modification}
\ee
where $V_{WS}$ is the total volume of the Wigner-Seitz cell,  and $E_{ETF}(A,Z)$ is the energy of a nucleus 
$(A,Z)$ in the vacuum defined by Eq.~(\ref{eq:ETF}).
We can also express $E_{ETF}(A,Z)$ from Eq.~(\ref{esurf}) as,
\be
E_{ETF}(A,Z)=\frac{\mathcal{E}_{sky}^{ETF}\left [\rho_{sat}(\delta),\rho_{sat}(\delta)\delta \right ]}{\rho_{sat}(\delta)} A  +E_S .
\label{energy_nucleus_vacuum}
\ee
Finally the total ETF energy can be decomposed as:
\bea
E_{ETF}^{tot}&=&\int_ 0^{R_{HS}}  \mathcal{E}_{sky}^{ETF}\left [\rho^{(0)},\rho^{(1)} \right] d^3r  \nonumber \\
&&\hspace{1cm}+ \int_{R_{HS}}^{R_{WS}} \mathcal{E}_{sky}^{ETF}\left [\rho^{(0)},\rho^{(1)} \right] d^3r  \nonumber  \\
&=& \mathcal{E}_{sky}^{ETF}\left [\rho_{sat},\rho_{sat}\delta_{sat}\right ] V_{HS} + 
\mathcal{E}_{sky}^{ETF}\left [\rho_g,\rho_g\delta_g\right ] \left (V_{WS}-V_{HS} \right ) \nonumber \\
 &&\hspace{1cm} + E_{S,m}.  
\label{emedium}
\eea
and, as shown in Ref.~\cite{pana,esym}, the total ETF energy is constituted of both a bulk  and a surface term.
In Eq.~(\ref{emedium}),  
$V_{HS}=A/\rho_{sat}$ the hard-sphere volume of the cluster, and $E_{S,m}$ represents 
a surface term since the bulk parts have been highlighted.

Using Eqs.~(\ref{total_energy_modification}), (\ref{energy_nucleus_vacuum}) and (\ref{emedium}),
we can express the total in-medium modification $\delta E_{m} = \delta E_B + \delta E_S$ as a bulk and a surface
term, with
\bea
\delta E_B =
 -A \frac{ \mathcal{E}_{sky}^{ETF}\left [\rho_g,\rho_g\delta_g\right ]}{\rho_{sat}(\delta)}
\; \; ; \; \;
\delta E_S =
E_{S,m}-E_S.
\label{deltaE}
\eea
Since $\delta E_S$ is 
proportional to two surface terms deduced from Eqs.~(\ref{energy_nucleus_vacuum}) and (\ref{emedium}),
we can expect
the following relation to hold: $\delta E_S = c_{s}A^{2/3}$,
where the parameter $c_s$ should have a weak dependence on $A$,  revealing the small effect of the curvature terms.
The validity of this decomposition will be explicitly tested below.

Finally the in-medium modified cluster energy, including both the bulk and the surface energy shift, is given by:
\bea
E_{m}(A,Z) &=& E_{ETF}(A,Z) + \delta E_{m} 
=  E_{B,m} + E_{S,m},
\eea
where
\bea
E_{B,m} =
\Big( \mathcal{E}_{sky}^{ETF}\left [\rho_{sat}(\delta),\rho_{sat}(\delta)\delta \right ] -
 \mathcal{E}_{sky}^{ETF}\left [\rho_g,\rho_g\delta_g\right ] \Big )\frac{A}{\rho_{sat}(\delta)} \nonumber \\
 \label{emedium_bulk}
\eea
and
\bea
E_{S,m}= E_S+\delta E_S.
 \label{emedium_fin}
\eea

In the next section, the medium modification of the bulk and surface energies are studied. In practice, a large set
of calculations is performed, varying the cluster size and isospin asymmetry over a large domain of $N$ and $Z$ 
covering the whole periodic table well beyond the neutron dripline.
Preliminary results with a simpler TF functional (zero order in $\hbar$) were already presented in ref.~\cite{pana,esym}.
As we show in the next sections, the inclusion of higher order terms slightly modifies the absolute values of the energy shifts, but does not modify the general trends reported in ref.~\cite{pana,esym}.

\subsection{Medium modifications of the bulk energy}

\begin{figure}[tb]
\includegraphics[angle=270,width=\linewidth,clip]{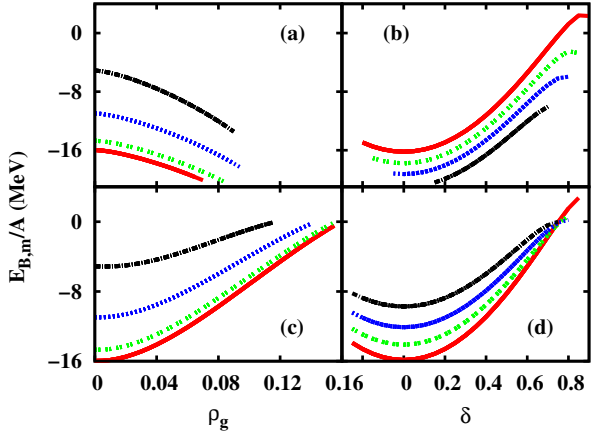}
\caption{(Color online) In-medium bulk energy $E_{B,m}/A$, defined by Eq.~(\ref{emedium_bulk}),
as a function of the gas density for a fixed bulk asymmetry (left side)
and as a function of the bulk asymmetry for a fixed gas density (right side). 
Upper panels: pure neutron gas ($\delta_g=1$). Lower panels:
gas asymmetry equal to the bulk asymmetry ($\delta_g=\delta$).
(a) and (c): $\delta=0.0$ (full red), $\delta=0.2$ (dashed green), $\delta=0.4$ (dotted blue), $\delta=0.6$ (dash-dotted black).
(b) and (d): $\rho_g=0.01$ (full red), $\rho_g=0.04$ (dashed green), $\rho_g=0.06$ (dotted blue), $\rho_g=0.08$ (dash-dotted black).
}
\label{fig:bulk}
\end{figure}

The in-medium bulk energy per nucleon $E_{B,m}/A$, defined by Eq.~(\ref{emedium_bulk}), and 
computed with the SLy4 interaction is displayed in Fig.~\ref{fig:bulk}
as a function of the gas density (left side) for different bulk asymmetries of the nucleus,
and as a function of the bulk asymmetry (right side) for different gas densities. 
Two representative cases are considered: a gas asymmetry equal to the cluster one
$\delta_g=\delta$ (lower panels)
and a pure neutron gas $\delta_g=1$ (upper panels). 
 
For very neutron rich clusters with $\delta\approx 1$, the case  $\delta_g=1$ is relevant 
both for the ground state of the neutron star inner crust, and for the most representative configurations 
of neutron rich matter at finite temperature. For nuclei close to isospin symmetry,
$\delta\approx 0$, the case $\delta_g=\delta$ corresponds to the most probable configurations
at finite temperature. 
In all cases, increasing gas density corresponds to physical situations at higher 
density and/or temperature.

Imposing the gas asymmetry to be strictly equal to the cluster asymmetry, 
amounts to disregard isospin effects (isospin fractionation) in the equilibrium.
In this case we recover the well known result that the cluster energy is 
reduced by the presence of the surrounding medium, leading to the 
 dissolution of clusters at the critical Mott density~\cite{typel,roepke}. 
The critical Mott density can be defined as the density corresponding to vanishing 
bulk binding, and is given by the ending point of each curve in Fig.~\ref{fig:bulk}(c). 
It is by construction the saturation density $\rho_{sat}(\delta)$ and we recover that it 
monotonically decreases with increasing cluster asymmetry~\cite{pana}.

In the case of stellar matter at $\beta$-equilibrium the fractionation effect 
cannot be neglected, and the gas is systematically more neutron-rich than 
the clusters. In particular, in the specific case of cold neutron star crust,
the uniform gas is uniquely constituted of neutrons~\cite{negele_vautherin}.
The limiting case $\delta_g=1$ is thus close to the physical condition of the 
low temperature stellar environment. 
In this case the trend with respect to the density (at fixed asymmetry $\delta$) is reversed. 
 
 The reduction of the in-medium bulk energy with respect to the density at
fixed $\delta$ is simple to understand:
the first term in Eq.~(\ref{emedium_bulk}) is constant at fixed $\delta$,
as well as the common factor $A/\rho_{sat}(\delta)$, while the second
term in Eq.~(\ref{emedium_bulk}) is increasing with the gas density at fixed
$\delta_g=1$.
While this effect seems a bit academic in panel (a), the consequence of the shift is
more interesting to comment in panel (b).
It is known that in the sequence of nuclei predicted in the crust of neutron 
stars~\cite{negele_vautherin}, as the density increases, the asymmetry
in the bulk of the nuclear clusters $\delta$ also increases.
This sequence can be understood in part from Fig.~\ref{fig:bulk}(b) since
as $\rho_g$ increases, the constant bulk energy path is going towards
more and more asymmetric clusters.
Taking the sequence of ground state nuclei predicted in the crust of neutron
stars~\cite{negele_vautherin}, the bulk energy departs from a quadratic 
behavior with respect of the bulk asymmetry $\delta$~\cite{pana} since
increasing the gas density shifts down the bulk energy, as shown in
Figs.~\ref{fig:bulk}(a) and (b).
This simple mechanism explains why clusters can survive in environment 
extremely neutron rich as neutron star crusts.  

It is however surprising that for the gas densities considered in Fig.~\ref{fig:bulk},
the medium modifications to the bulk energy remain mostly quadratic with
respect to $\delta$ at fixed $\rho_g$.
Non-quadraticities in $\delta$ are  only observed 
 for $\delta \geq 0.6$,  with or without gas. (right part of Fig.~\ref{fig:bulk}).
The quadratic dependence of the bulk energy with respect to $\delta$ is therefore
a robust prediction which goes beyond the case of isolated nuclei and can be generalized
to dilute nuclei to a large extent.

\subsection{Medium modification of the surface energy}

\begin{figure}[tb]
\includegraphics[angle=270,width=\linewidth,clip]{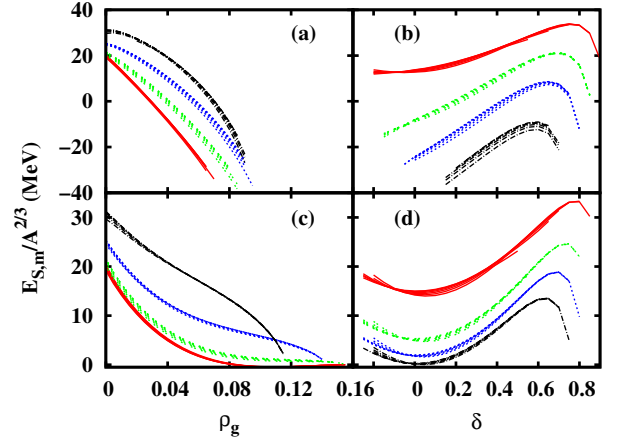}
\caption{(Color online) In-medium surface energy (see Eq.~(\ref{emedium_fin})) 
as a function of the gas density for a fixed bulk asymmetry (left part)
and as a function of the bulk asymmetry for a fixed gas density (right part) (see text).
Upper panels: pure neutron gas. Lower panels:
gas asymmetry equal to the bulk asymmetry.
(a) and (c): $\delta=0.0$ (full red), $\delta=0.2$ (dashed green), $\delta=0.4$ (dotted blue), $\delta=0.6$ (dash-dotted black).
(b) and (d): $\rho_g=0.01$ (full red), $\rho_g=0.04$ (dashed green), $\rho_g=0.06$ (dotted blue), $\rho_g=0.08$ (dashed-dotted black).
}
\label{fig:surf}
\end{figure}

Fig.~\ref{fig:surf} illustrates the surface tension, defined as the scaled in-medium surface
energy  $E_{S,m}/A^{2/3}$, cf Eq.~(\ref{emedium_fin}), as a function of the gas density $\rho_g$
and of the bulk asymmetry $\delta$  for the same gas compositions as for Figure  \ref{fig:bulk}. 
The almost perfect scaling with $A^{2/3}$ shows that indeed the in-medium modification of the binding 
energy is mainly a surface effect.
There are only few cases where the curves acquire a finite width, reflecting a small contributions from
curvature terms: In Fig.~\ref{fig:surf}(a) where $\delta_g=1$ and for the most neutron rich clusters
(black curves), and in Fig.~\ref{fig:surf}(c) where $\delta_g=\delta$ and here also for the most neutron 
rich curves (for e.g. black curves at $\rho_g=0$).
The curvature terms have been discussed in Sec.~\ref{surf2}, and are observed here to be maximal
in the most asymmetric clusters as the gas density increases.

Neglecting fractionation effects in Fig.~\ref{fig:surf}(c),  the surface 
energy is reduced as the gas density increases and whatever the cluster asymmetry.
It vanishes at the corresponding saturation density $\rho_{sat}(\delta)$,  showing again the 
dissolution of clusters in the dense medium. 
In Fig.~\ref{fig:surf}(d), the dependence of the surface energy with $\delta$ is
mostly quadratic, even for the largest densities considered here.
The quadratic behavior of the surface energy is well satisfied up to $\delta\geq 0.6$,
as in the case of the bulk energy.
  
It is quite surprising to find in the case of pure neutron gas, Figs.~\ref{fig:surf}(a) and (b),
that the surface energy not only decreases as the gas density increases, but can even
become negative.
This can be understood from the fact that 
the surface energy as defined by Eq.~(\ref{emedium}) represents the interface contribution
between the cluster and the gas. 
At finite gas density, this interface energy contains contributions from both the cluster and the gas.
The contribution of the pure neutron gas to the interface region is largely negative, since the
interface region is more symmetric than the gas.
The negative contribution of the gas dominates as the gas density increases, leading to negative surface energy
as shown in Fig.~\ref{fig:surf}(a).
This effect is lowered when the cluster is more neutron rich, see Fig.~\ref{fig:surf}(b).

It should also be remarked that the density as which the surface energy becomes negative 
increases as the bulk asymmetry increases.
Since the ground state configurations predicted for the crust of neutron stars~\cite{negele_vautherin},
have increasing $\delta$ for increasing $\rho_g$, these configurations always correspond to systems
where the surface energy is positive\cite{pana,esym}.
Concerning the dependence of the surface energy on $\delta$ in Fig.~\ref{fig:surf}(b),
we can see a very different behavior compared with the previous cases:
the  quadratic approximation in $\delta$ is completely lost
due to the contribution of the gas, which is
not quadratic in $\delta$, but in $\delta_g$.

\subsection{Dependence on the effective interaction}

\begin{table}
\caption{Bulk nuclear properties for the different Skyrme interactions
examined in this paper.}
\begin{ruledtabular}
\begin{tabular}{lrrrrr}
NN-potential & $\rho_{sat}$   & $K$  & $L_{sym}$   & $K_{sym}$ & 
$E_{sym}$ \\
$~$          & (fm$^{-3}$)& (MeV) & (MeV) & (MeV) & (MeV)     \\
\hline\noalign{\smallskip}
SLY4         & 0.159    & 230.0   & 46.0  & -119.8 & 
  32.00     \\
SGI          & 0.154     & 261.8   & 63.9  & -52.0 & 
   28.33      \\
SkI3         & 0.158    & 258.2  & 100.5 & 73.0  & 
    34.83    \\
LNS          & 0.175    & 210.8 & 61.5  & -127.4 & 
    33.43     \\  
\end{tabular}
\end{ruledtabular}
\label{table:NNparam}
\end{table}

\begin{figure}[tb]
\includegraphics[angle=270,width=\linewidth,clip]{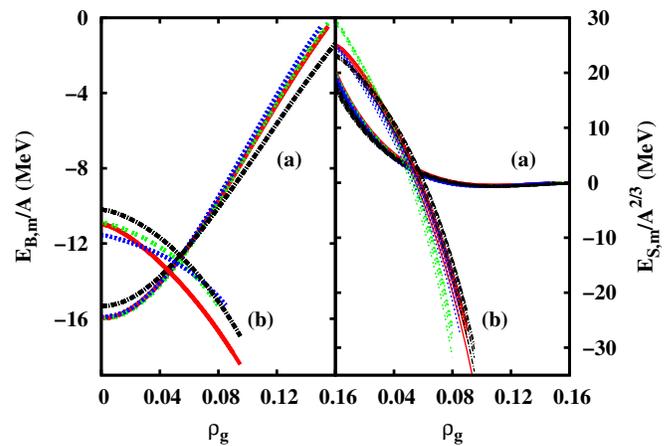}
\caption{(Color online) In-medium bulk (left) and surface (right) energy as a function of the gas density (see text).
(a): clusters with bulk asymmetry $\delta=0$ immersed in a symmetric gas.
(b): clusters with  $\delta=0.3$ immersed in a pure neutron gas. 
Different models are considered:
 Sly4 ~\cite{sly4} (full red), SkI3 ~\cite{ski3} (dashed green), SGI~\cite{sgi} (dotted blue), LNS ~\cite{lns} (dash-dotted black).}
\label{fig:eff_int}
\end{figure}

In this section, we show that the qualitative behaviors that we have discussed in this paper 
are not modified if a different Skyrme interaction is employed. 
In particular, the positive sign of the surface symmetry energy that we have discussed in Sec.~\ref{surf1}
does not depend on the particular effective interaction. 
However the quantitative values of the clusters bulk and surface energies obviously depend on the 
effective interaction parameters, and for a realistic treatment of the stellar matter equation of state 
it is very important to consistently treat withing the same effective interaction both the 
cluster and the free gas~\cite{pana,esym}.

To study how the in-medium effects depend on the model, we represent in Fig.~\ref{fig:eff_int} two 
representative situations of a symmetric nucleus in a symmetric gas, and a neutron rich nucleus in a 
pure neutron gas, with different Skyrme models. We have chosen these specific interactions in order to 
span the present uncertainties in the  bulk parameters.
These latter are reported in Table  \ref{table:NNparam}.

We can see that the qualitative behavior of the different models is the same.
A more complete study of the effective interactions parameter space is needed to reach
sound conclusions on the quantitative model dependence, but from the representative chosen
interactions we can dress some tentative partial interpretation.
The differences in bulk energy directly reflect the uncertainties in present models of the 
bulk properties of matter. 
These uncertainties are very small in the isoscalar part, and the curves of the bulk symmetric systems, 
see curves labelled (a) on the left panel, are indistinguishable except LNS (dashed-dotted black lines). 
The LNS Skyrme model is known to have a saturation density larger that the expected one, see Tab.~\ref{table:NNparam},
which is reflected in the fact that at $\rho_g=0$~fm$^{-3}$, the LNS bulk energy is different from the others.
It also leads to slightly reduced in-medium modification, as observed in Fig.~\ref{fig:eff_int}.
Concerning medium modifications to the bulk energy in the neutron rich system, see curves labelled (b) on the left panel, 
it is observed that SLy4 Skyrme interaction (full red) leads to slightly more important binding energy shift. 
This is due to a non-trivial interplay of slightly different values of $E_{sym}$, $L_{sym}$, $K_{sym}$. 
Concerning the surface energies, the behavior appears very stable. 
The only exception is the gas density behavior of the neutron rich system, curves (b) in the right panel,
calculated with SkI3 (dashed green). 
This steep in medium modification is probably due to the very stiff isovector properties of this effective interaction. 

To conclude, we can see that, independent of the model, the in-medium modification are not negligible 
and should be accounted for in a realistic equation of state.
Due to the simple expression  (\ref{emedium}), these corrections can be 
tabulated as a function of $(A,I,\rho_{g,n},\rho_{g,p})$ and straightforwardly 
introduced in the equation of state calculations~\cite{pana,esym} 
as a modification of the cluster energy functional with no extra computational cost.

\section{Conclusions}

In this paper we have considered a simple analytical modelling of the nuclear density profiles allowing 
to calculate  nuclear binding energies within the Extended 
Thomas Fermi approximation at the second order in $\hbar$.
Through a comparison to HF calculations for some representative nuclei,
we have shown that a simple Fermi-Dirac profile is sufficiently flexible to reach a precision in the energy of 
the order of a 100-200 keV/nucleon, and the widening of the variational space 
considering FGD trial densities 
does not introduce any sizeable improvement of the predictive power of the model.

Two different applications of the model were presented.
The first one concerns the definition of the bulk and surface part of the symmetry energy 
of finite nuclei, which is important for the extraction of equation of state parameters for astrophysical applications.
We have shown that the variational character of the ETF formalism suggests that the bulk part of the nuclear energy
depends on the central bulk asymmetry $\delta$ rather than on the global asymmetry of the nucleus $I$ which
is usually considered in LDM.
This statement, which is confirmed by a detailed comparison to HF calculations, implies that the
surface symmetry energy contributes positively to the total symmetry energy of the nucleus.
The choice of the global asymmetry parameter $I$ considered in LDM, while not consistent with ETF,
explains the ambiguities reported in the literature concerning the sign of the surface symmetry energy.

The second application concern the evaluation of the in-medium energy shift which is experienced by a nucleus
immersed in the gas of its continuum states, as it is the case in supernova matter and in the inner crust
of (proto)-neutron stars. We have shown that the presence of an external gas induces both a bulk and a surface
energy shift, which depend in a highly complex and non-linear way on the asymmetry of the cluster, and the asymmetry 
and density of the gas. The absolute values of these energy shift can be comparable or higher than the nuclear 
binding energy, meaning that the coexistence of nuclei and free particles in stellar matter cannot be modelized 
as a mixture of non-interacting nuclear species as it is done in the current models of stellar equation of state~\cite{ esym,hempel,japanese}.
 
\begin{acknowledgments}
This work has been partially funded by the SN2NS project ANR-10-BLAN-0503 and it has
been supported by New-Compstar, COST Action MP1304.
\end{acknowledgments}

\newpage

\appendix

\section{Coefficients of the Skyrme functional} \label{coefficients}

Here we write the coefficients of the Skyrme functional as given by Ref.~\cite{Bender2003a},
\begin{eqnarray}
C_0^\rho &=& \frac{3}{8}t_0 + \frac{3}{48}t_3 \rho^{(0)\alpha}(r) \\
C_1^\rho &=& -\frac{1}{8}t_0(2x_0+1) - \frac{1}{48}t_3(2x_3+1) \rho^{(0)\alpha}(r) \\
C_0^{\Delta\rho} &=& \frac{1}{64}[-9t_1+t_2(4x_2+5)]\\
C_1^{\Delta\rho} &=& \frac{1}{64}[3t_1(2x_1+1)+t_2(2x_2+1)]\\
C_0^\tau &=&\frac{1}{16}[3t_1+t_2(4x_2+5)]\\
C_1^\tau &=&\frac{1}{16}[-t_1(2x_1+1)+t_2(2x_2+1)]\\
C_0^{\nabla J} &=&-\frac{3}{4} W_0\\
C_1^{\nabla J} &=&-\frac{1}{4} W_0\\
C_0^J &=&-\frac{1}{16}[t_1(2x_1-1)+t_2(2x_2+1)]\\
C_1^J &=&-\frac{1}{16}[-t_1+t_2]
\end{eqnarray}

\section{Analytical density profile for symmetric nuclei in spherical symmetry} \label{analytic}

Following the derivation of Ref.~\cite{krivine}.
our starting point is the one-dimensional Euler equation given by Eq.~(\ref{eq:Euler_1D}),
\begin{equation}
\lambda=\frac{dh}{d\rho}+   C^\nabla  \left(  {\nabla}\rho\right)^2
+C^\Delta \Delta\rho
, \label{start}
\end{equation}
In the limit of very large $r$,  this equation simplifies to:
\bea
\lambda = 
\frac{1}{36} \frac{\hbar^2}{2m}
	\left[
			\left(    \frac{ \rho' }{\rho}    \right)^2
		-2  	\frac{\rho''}{\rho}
	\right]
,
\label{eq_lambda_sym_infinite_x+infty}
\eea
which gives
\bea
\rho(r) \propto \e^{-r/a_{out}}
&
\textrm{\;\; with\;\; }
&
a_{out} = 
\sqrt{  
- \frac{\hbar^2}{2m} \frac{1}{36\lambda}  
}
\label{eq_sym_infinite_rho_x+infty}
\eea
The value of $\lambda$ is obtained by considering the bulk limit of Eq.~(\ref{start}). In this limit we have:
\begin{align}
\lambda= \left. \frac{\partial \mathcal{E}_{sky}^{ETF}}{\partial \rho} \right|_{\rho_{sat}}
= \left. \frac{\mathcal{E}_{sky}^{ETF}}{\rho} \right|_{\rho_{sat}} \equiv \lambda_{sat}
,
\label{eq_,ym_lambda_nucl_matter}
\end{align}
where we can recognize $\lambda_{sat}$ 
as the chemical potential of symmetric nuclear matter at saturation.

Close to the bulk limit, that is for $r\to 0$ and $R_\nu\to\infty$,   linearizing Eq.~(\ref{start}) 
introducing $\rho(r)=\rho_{sat} +\delta \rho$ gives
\be
\lambda_{sat} = 
 \frac{dh}{d\rho}  (\rho_{sat})
+  \frac{d^2 h}{d\rho^2}  (\rho_{sat})  \delta\rho
+ C^\Delta(\rho_{sat})				\delta\rho''
,
\label{eq_lambda_sym_infinite_x-infty}
\ee
where  we have 
defined $f_{sat}=f(\rho_{sat})=1+\kappa\rho_{sat}$. 
Solving equation 
 (\ref{eq_lambda_sym_infinite_x-infty}) leads to
\be
\delta\rho(r) \propto \e^{(r-R_\nu)/a_{in}} \label{eq_sym_infinite_rho_x-infty}
\ee
with
\bea
\frac{K_{sat}}{9}a_{in}^2 &=& 
		 \frac{\hbar^2}{2m}
		 \frac{1}{3}	
\left[	\frac{1}{6} - \frac{7}{3}\kappa\rho_{sat}   +  \frac{\kappa\rho_{sat}}{2f_{sat}} \right] \\ \nonumber
				&-&C_0^{\nabla J}B_J \frac{\rho_{sat}^2}{f_{sat}} 
				+ 2C_0^{\Delta\rho} \rho_{sat}
		,
\label{eq:anu}
\eea
where  
 $K_{sat}=9\rho_{sat}\partial^2\mathcal{E}_{sky}^{ETF}/\partial\rho^2|_{\rho=\rho_{sat}}$
is the nuclear matter incompressibility.
To achieve the two  asymptotic behaviors, the density profile can be represented 
as a generalized Fermi function (GFD) $\rho = \rho_{GFD}= \rho_{sat} F_\nu$ with:
\be
 F_\nu(r) = \left( 1 + \e^{(r-R_\nu)/a_\nu} \right)^{-\nu}
\label{eq_sym_infinite_density_gen_F}
\ee
Comparing equations  (\ref{eq_sym_infinite_rho_x+infty}) and  (\ref{eq_sym_infinite_rho_x-infty}) with  (\ref{eq_sym_infinite_density_gen_F}), we have 
\be
a_\nu = a_{in} \;\; ;\;\;  
\nu  = \frac{a_{in}}{a_{out}}=\frac{6a_\nu}{\hbar} \sqrt{-  2m\lambda_{sat} },
\label{eq:nu}
\ee

The link between the parameter $R_\nu$ and the particle number is deduced from the leptodermous series 
developement of $a_\nu/R_\nu$ of the integral giving the particle number:
\be
A = 4\pi \int_0^\infty \mathrm dr \rho_{GFD}(r)r^2   
\label{eq_sym_3D_A}
.
\ee
The moments $I^m_{\nu}$  of the GFD ansatz have been calculated in Ref.~\cite{krivine_math}, as, 
\bea
I^m_{\nu} &=&
\int_0^{+\infty} \mathrm dr F_{\nu}(r)r^m \\ \nonumber
&\simeq&
\frac{R_\nu^{m+1}}{m+1}
	\left[
		1 + (m+1)
				\sum_{n=0}^m \binom{m}{n}  \eta^{(n)}_\nu  \left( \frac{a_\nu}{R_\nu} \right)^{n+1}
	\right]
,
\label{eq_annex_Imunu}
\eea
where:  
\bea
\eta^{(n)}_{\nu} =
(-1)^n
\int_0^\infty \mathrm du
\left[
	\frac{1+ (-1)^n \e^{-\nu u}}{\left(1+\e^{-u}\right)^\nu} 
	-1
\right]
u^n
,
\label{eq_annex_gen_eta}
\eea
Replacing in Eq.~(\ref{eq_sym_3D_A}) gives:
 
\be
A = 
\frac{4}{3} \pi \rho_{sat} R_\nu^3 	\left[
										1 +
										3 \eta^{(0)}_\nu \frac{a_\nu}{R_\nu} +
										6 \eta^{(1)}_\nu \left( \frac{a_\nu}{R_\nu} \right)^2 +
										3 \eta^{(2)}_\nu \left( \frac{a_\nu}{R_\nu} \right)^3
								\right]  \nonumber
.
\ee
If $a_\nu \ll R_\nu$ this expression can be inverted giving at  third order in the leptodermous expansion:
\bea
\frac{R_\nu}{R_{HS}} &\simeq&
		1 
		-
		\eta^{(0)}_\nu \frac{a}{R_{HS}} 
		+
		\left[ 
			\left(  \eta^{(0)}_\nu  \right)^2
			-2\eta^{(1)}_\nu 
		\right] 
			\left( \frac{a}{R_{HS}} \right)^2  \nonumber \\
		&-&
		\left[
				\frac{2}{3} \left( \eta^{(0)}_\nu \right) ^3
				- 2 \eta^{(0)}_\nu \eta^{(1)}_\nu
				+ \eta^{(2)}_\nu 
		\right]
			\left( \frac{a}{R_{HS}} \right)^3
,
\label{eq_sym_3D_R_nu}
\eea
where $R_{HS} = (3A/4\pi\rho_{sat})^{1/3}$   is the equivalent homogeneous sphere radius.

In conclusion, the parameters of the GFD ansatz given by the parametric form~(\ref{eq:GFD}) and
(\ref{eq_sym_infinite_rho_x-infty}) can be determined 
in the following way: $\rho_{sat}$ is the saturation density of nuclear matter, 
$a_\nu$ is given by Eq.~(\ref{eq:anu}), $\nu$ is given by Eq.~(\ref{eq:nu}) and $R_\nu$ is given by Eq.~(\ref{eq_sym_3D_R_nu}).

\section{Inclusion of the  spin-orbit current for symmetric nuclei in spherical symmetry} \label{bj}

The nuclear interactions considered in this work neglect the contribution of the spin-orbit current $J^{(t)2}$ in the 
functional~(\ref{eq:functionalt}).
In this section, we give the corrections to be applied to the Euler equation without neglecting the spin-orbit current.

In symmetric nuclei, the spin-orbit potential $W$~(\ref{eq:sopotential}) reduces to the simpler form,
\begin{equation}
W =C_0^J J-C_0^{\nabla J} \nabla\rho .
\label{eq:sopot}
\end{equation}
Injecting Eq.~(\ref{eq:sopot}) into Eq.~(\ref{eq:socurrent}) gives
\begin{equation}
J=B_J(\rho)\frac{\rho \nabla \rho}{f} ,
\end{equation}
with
\begin{equation}
B_J(\rho) = \frac{2m}{\hbar^2}\frac{C_0^{\nabla J}}{1+\frac{2m}{\hbar^2}C_0^J \frac{\rho}{f}} .
\label{eq:BJ}
\end{equation}
Setting $C_0^J=0$ in Eq.~(\ref{eq:BJ}) allows to recover the definition of the constant $B_J$
used in this work.
Now the coefficient $B_J$~(\ref{eq:BJ})  is a function of the density, and it will modify the
Euler-Lagrange equation~(\ref{eq:Euler}).

The correction to Eq.~(\ref{eq:Euler_1D}) induced by the spin-orbit current is given by the
modification of only two terms:
\begin{eqnarray}
C^{\nabla}(\rho) &\rightarrow& C^{\nabla}(\rho)+\frac{1}{2}C_0^{\nabla J}\frac{d B_J(\rho)}{d \rho} \frac{\rho}{f} \\
\frac{d h}{d\rho} &\rightarrow&\frac{d h}{d\rho} -C_0^{J2}B_J(\rho)\frac{d B_J(\rho)}{d\rho}\rho .
\end{eqnarray}


\end{document}